\documentclass{article}

\usepackage[letterpaper]{geometry}
\usepackage[table]{xcolor}
\usepackage{graphicx}
\usepackage[textwidth=8em,textsize=small]{todonotes}
\usepackage{amsmath}
\usepackage{amssymb}
\usepackage{amsthm}
\usepackage{amsfonts}   
\usepackage{authblk}
\usepackage{url}
\usepackage{hyperref}
\usepackage{cleveref}
\usepackage{nicefrac}       
\usepackage{microtype}      
\usepackage{lipsum}
\usepackage{graphicx}
\usepackage{natbib}
\usepackage[shortlabels]{enumitem}
\usepackage{mathtools}
\usepackage{stmaryrd} 
\usepackage[font=small,labelfont=bf]{caption}
\usepackage{subcaption}
\usepackage{appendix}
\usepackage{soul}
\usepackage[anythingbreaks]{breakurl}

\usepackage{breakurl}
\usepackage{booktabs}       

\usepackage{natbib}
\setcitestyle{authoryear,open={(},close={)}}

\newcommand{\bs}[1]{\boldsymbol{#1}}

\usepackage{pifont}

\newtheorem*{theorem*}{Theorem}
\newtheorem*{lemma*}{Lemma}

\newcommand\blinded[1]{\emph{blinded for review}}

\DeclareMathOperator*{\argmax}{arg\,max}

\begin{document}

\title{\textbf{Estimating soil organic carbon storage potential and approximating optimal management policies} }

\author[1]{Jacob V. Spertus}
\author[2,3]{Eric W. Slessarev}
\author[3]{Whendee L. Silver}
\author[1]{Philip B. Stark}

\small{
\affil[1]{Department of Statistics, University of California, Berkeley}
\affil[2]{Department of Ecology and Evolutionary Biology, Yale University}
\affil[3]{Environmental Science, Policy, and Management, University of California, Berkeley}
}

\date{September 10th, 2026}

\maketitle

\begin{abstract}
    The impact of a management intervention on the soil organic carbon (SOC) stored in a given volume of soil may be moderated by observable features that proxy that soil’s SOC storage potential under that intervention. 
    To maximize total SOC storage, interventions should be targeted to soils with the highest responses and lowest costs to intervening. 
    We summarize key sources of uncertainty and present a causal framework for estimating SOC storage potentials and optimal management policies. 
    The method models SOC measurements as functions of covariates within each treatment arm, using the fitted models to estimate SOC storage potential for each plot and finding the policy that maximizes the average of those estimates. 
    The modeling can use linear regression or other machine learning algorithms to learn relationships between features and SOC storage. 
    We demonstrate its use in a study of compost amendments applied to California rangelands, finding that storage potential is moderated by baseline SOC and that optimizing the policy could provide a slight gain over uniform application of the intervention with the largest average treatment effect estimate. 
    We evaluate the methods further in simulated field experiments. 
    Simple estimates fared better than machine learning, except in unrealistically large experiments. 
    We conclude by discussing recommendations for practice, baseline SOC moderation, extensions to observational studies, and broader policy uncertainties.

    \vspace{0.5em}
\end{abstract}

\section{Introduction}

Increasing soil organic carbon (SOC) can improve various aspects of soil health---including nutrient and water retention, fertility, and erosion resistance---and remove atmospheric carbon dioxide (CO$_2$) when the amount stored is additional after accounting for exogenous C inputs and emission of non-CO$_2$ greenhouse gases \citep{lal2004, bossio2020role}. To support efforts to restore SOC, investigators must evaluate the effects of management changes on SOC and, ideally, tailor management policies to maximize storage.

More knowledge of soil properties might increase the efficiency of policies aimed at increasing SOC. 
For instance, the amount of SOC a particular volume of soil can store after a given intervention – its SOC \emph{storage potential} – is unknown. 
Storage potential determines (a) how policy should prioritize sites for intervention and (b) the limits of what intervention can accomplish, including the amount of CO$_2$ stored, the rate of drawdown, and the longevity of the storage. 
It is an inherently causal concept, comparing the effects of two or more courses of action on the same soil.

\begin{figure}
    \centering
    \includegraphics[width = \textwidth]{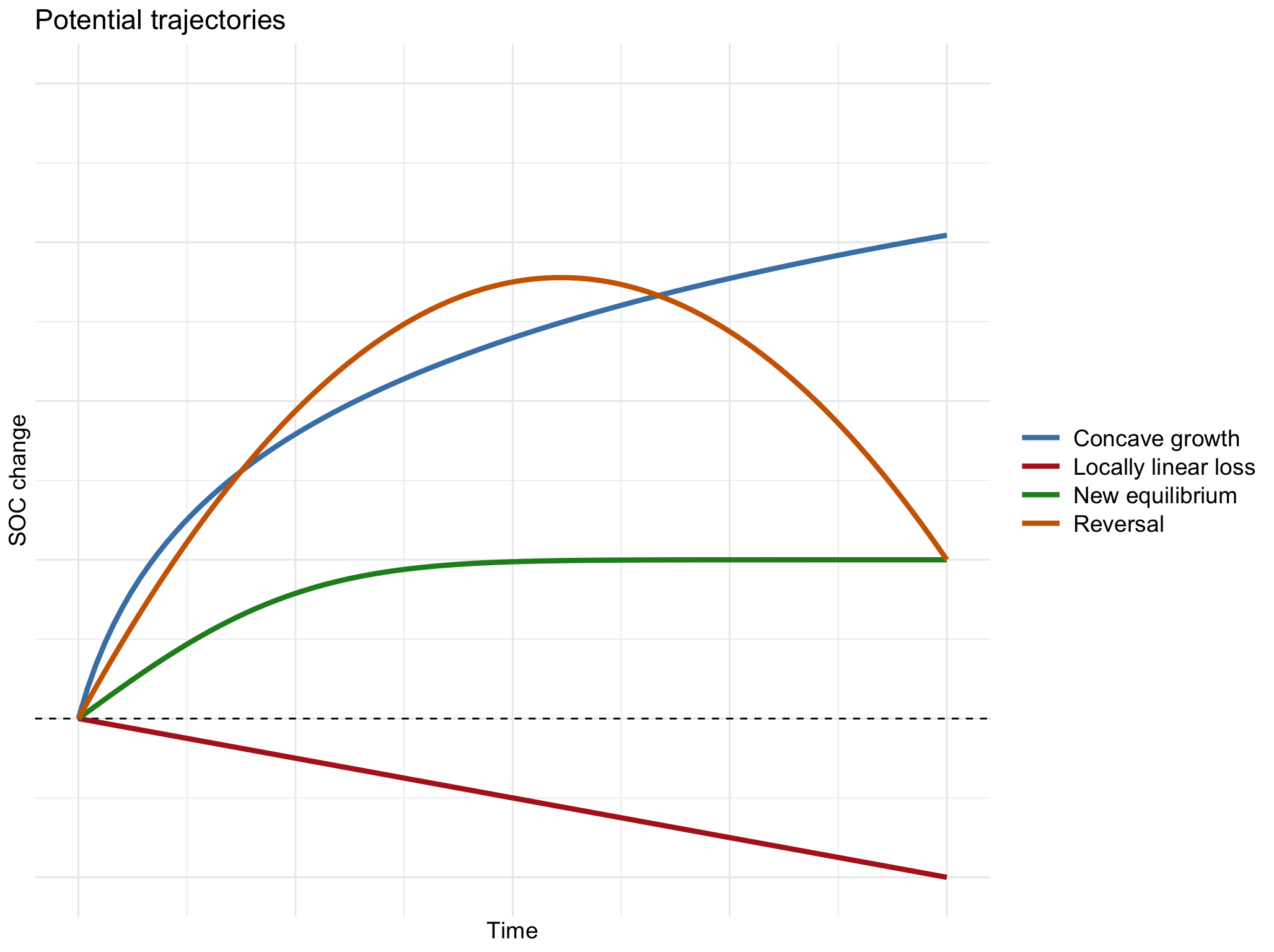}
    \caption[Potential soil carbon trajectories over time for different treatments]{Four idealized population-level potential trajectories (colored lines) of additional SOC sequestered (y-axis) over time (x-axis). From the point a decision is implemented, each course of action leads to a different trajectory of total SOC stored across the population. Note that these trajectories do not represent short-term temporal variation, which can be substantial \citep{wuest2024temporal}. A policy-maker with access to all potential trajectories and a target timeline could make optimal decisions under budgetary constraints. By definition, potential trajectories are equal at baseline (left most point on x-axis). The dashed line represents baseline SOC. Under the green trajectory, SOC approaches a new equilibrium after the change has occurred, as theorized in \citet{stewart2007soil}. Under the blue and red trajectories, equilibrium is not achieved by the end of the time span plotted. The red trajectory displays a linear loss of SOC, as found, for example, in the study of \citet{sanford2012soil}. Under the orange trajectory, there is a reversal, a major concern of policies aiming to sequester SOC as a negative emissions strategy \citep{smith2005overview, thamo2016challenges}. 
    Potential reversals necessitate additional actions to maintain or increase C stocks in the future.}
    \label{fig:potential_trajectories}
\end{figure}

To illustrate, suppose a policy-maker is tasked with deciding whether to pay for a policy implementing no-till agriculture across all registered cropland in the US Corn Belt (the target \emph{population}) in 2030. 
Ideally, they would know the total amount of SOC in the top 1~m of soil on each farm in (say) 10 years, both if that farm adopted no-till agriculture for that period and if it carried on as usual, with bi-annual tillage. 
These quantities, only one of which can ever be observed on a given plot, are called \emph{potential outcomes}\footnote{
Potential outcomes were first described by Jerzy Neyman in his seminal 1923 paper on agricultural experiments \citep{neyman1923sur}. \citet{hurlbert1984pseudoreplication} provides a non-technical overview of terminology and issues in experimental design. \citet{imbens2015causal} is a comprehensive reference on causal inference with potential outcomes.}. 
As a function of time, potential outcomes are called \emph{potential trajectories} (\Cref{fig:potential_trajectories}). 
If the policymaker knew all the potential trajectories within a population of farms, they could tailor incentives to maximize total storage given the specific timing, costs, and impacts of each intervention on each farm.

SOC storage potential is a key component of SOC \emph{sequestration potential}, which incorporates changes in SOC storage, accounting for alternative fates of exogenous C inputs and for non-CO$_2$ greenhouse gas emissions. 
Sequestration rates are then compared among multiple courses of action. 
In a binary comparison, a management change is typically compared to some business-as-usual baseline that serves as control. 
\emph{Life cycle assessment} (e.g., \citet{delonge2013lifecycle}) is a crucial tool to estimate C inputs and uncertainties, allowing investigators to map between storage and sequestration. This paper focuses on the storage aspect, describing a formal method to estimate storage-optimal policies using empirical proxies of SOC storage potential.

\subsection{Classifying management interventions}

Management interventions may be conceptualized as policy changes or physical changes, and verification is required to ensure a physical management change actually occurs after a policy change. Furthermore, interventions could involve a single point in time (e.g., bolus of an input) or a pattern continuing over time (e.g., repeated application of inputs). The pattern should be clearly defined when designing and interpreting studies (e.g., researchers should disambiguate whether ‘adopting conservation tillage’ entails a switch at baseline or sustained maintenance of that strategy). Interventions could be fixed in advance, but in real-world management they are typically adjusted in light of changing soil, ecosystem, and socioeconomic conditions, resulting in a \emph{dynamic treatment regime} that can be modeled via reinforcement learning \citep{qian2011performance}. Dynamic management is beyond the scope of this paper: we assume hereafter that the intervention is fully defined at baseline either as a point-in-time or a pattern over time.

Interventions can be conceptualized as quantities, categories, or both. 
Quantitative interventions for SOC sequestration include amendments like compost, manure, fertilizer, or biochar application, which are readily summarized in terms of the amount applied (in total, per hectare, and/or per unit time). 
Categorical interventions include tillage practices, adaptive multi-paddock (AMP) grazing, and other broad land-use changes. 
A study that compares various intensities of compost and various intensities of chemical fertilizer contains both quantitative (intensity) and categorical (input type) interventions.

\subsection{Mechanisms governing SOC storage}

Interventions change SOC levels by changing the amount of C input to the soil and by changing the biological, chemical, and physical conditions that govern SOC retention in soil. 
Broadly speaking, SOC storage occurs when C inputs exceed losses from decomposition, which varies as a function of the chemical and physical characteristics of the organic material, environmental factors, and soil properties. When C inputs and losses are balanced, SOC is at equilibrium \citep{chenu2019increasing}. 
If inputs or conditions change, SOC tends toward a new equilibrium over time.

The baseline concentration of SOC in a given soil may influence the decomposition rate, and hence the soil’s capacity to store new C after an intervention in the absence of other changes. Baseline SOC concentrations may moderate SOC accrual via a variety of mechanisms. The capacity of soils to retain SOC is largely influenced by the availability and reactivity of soil mineral surfaces and the propensity for aggregation, both of which interact with organic matter and slow its decomposition. 
Because the amount of mineral surface area in a given volume of soil is finite, the amount of mineral associated organic carbon (MAOC) that a soil can store may saturate \citep{hassink1997capacity, georgiou2025soil}.  If the saturation hypothesis is true, management interventions would be expected to have less impact on fields near their effective capacity. This implies that baseline SOC levels would be a key moderator of SOC storage under a given set of environmental and mineralogical conditions. In practice, few studies consider the full depth profile available for C storage; even if surface soils slow or cease to sequester SOC, there may be gains lower in the profile where mineral content and SOC residence times are often higher \citep{schrumpf2013storage}.

Even if mineral surfaces do not saturate, microbial activity may scale non-linearly with SOC \citep{craig2021biological}. In this case, decomposition rates could depend on SOC levels, also implying baseline moderation. Because baseline moderation can be caused by multiple mechanisms, it cannot be used to directly test individual mechanisms (e.g., baseline moderation does not imply MAOC saturation). Quantifying the strength of baseline moderation is nonetheless important to provide empirical constraints on the amount of SOC that a given intervention might sequester in a given soil.

\subsection{Policy optimization and its pitfalls}

Covariates that proxy SOC storage potential—baseline SOC, baseline MAOC, soil type, land-use history, normalized difference vegetation index (NDVI), or other features—could play two key roles in improving scientific research and policy decisions. First, they can be used to form model-assisted estimates of \emph{average treatment effects}, which quantify the C storage of ‘blanket’ policies that apply the same management to every site. Average treatment effects are often poorly constrained because of the high signal-to-noise ratio in SOC measurement \citep{viscarrarossel2024how, stanley2023valid}, but are critical to C accounting and mitigation projections (e.g., \citet{iea2023netzero}). Second, proxies of SOC storage potential can be used to help target interventions and design more efficient management policies.

Our methods aim to maximize the total SOC stored across a population. 
This goal is utilitarian, normatively assuming that the best action is the one with the highest total SOC storage. It may be more ethical or practical to maximize another population-level welfare function, like minimum storage across plots. 
Targeting the total is partially justified by the role of SOC storage in atmospheric CO$_2$ removal, which reduces to the sum of sequestration on individual plots (i.e., the sum of SOC storage minus the sum of C costs). Any given policy choice will affect many other outcomes that are \emph{externalities}\footnote{
 To name a few: the health and productivity of soil, the flux of greenhouse gasses besides CO$_2$ (e.g., methane and nitrous oxide), the financial well-being of farmers and rural communities, the availability of habitat for wildlife, the health of people and livestock consuming the produce, the continuity of relationships between people and land, the purity of water, the beauty of landscapes, etc.
} in our development. 
From a bird’s eye view, this paper essentially describes a way to centrally plan agriculture with a decision algorithm. 
The ethics and efficacy of such programs are contested, often contrasted against decentralized stewardship styles characteristic of indigenous agroecology or smallholder farming (e.g., \citet{scott1999seeing, berry2004unsettling}). 
Our aim is to support efficient SOC research and policy, which requires evaluating both the potentials and the limitations of quantification and decision algorithms. 
With that in mind, we first provide a comprehensive survey of uncertainties. 
We then propose and evaluate a method to estimate SOC storage potentials and optimal management policies.

\section{Sources of uncertainty}
\label{sec:uncertainties}

\subsection{Core-level uncertainties}
To measure SOC stocks, investigators typically measure subsamples of soil cores for SOC concentration (\%SOC) and use the whole volume or mass for dry bulk density (BD) or equivalent soil mass (ESM; see next section). 
At the core level, uncertainties in \%SOC arise from laboratory sample preparation protocols, human error, subsampling variability, and instrument error, which we collectively call \emph{assay variability}. 
For instance, the presence and specific handling of particulate organic matter like plant fragments can cause considerable variation between sub-sampled assays from the same soil core \citep{wuest2024temporal}. 
Assay variability can be addressed through careful laboratory work, including the use of precise instruments (e.g., elemental analyzers rather than loss on ignition), standard reference material, and analytical replicates to control and estimate subsampling and instrument error. 
Determining the volume of soil via BD is also subject to core-level uncertainties, including compression of soils during sampling, the presence of coarse fragments (e.g., gravel) in BD cores, and residual moisture. 
The use of ESM is less sensitive to the volume of soil sampled but still incorporates uncertainty within and across samples due to coarse fragment fractions and accurate sampling to depth. 
When MAOC is measured, the soil fractionation process requires a specific range of shaking or ultrasonic intensities to release the fine fraction without breaking up the coarse fraction and thereby contaminating the MAOC measurement with POC \citep{amelung1999minimisation, six2024six}.

\subsection{Plot-level uncertainties}

Plot-level uncertainties are often more substantial than core-level \citep{stanley2023valid}. The main source of plot-level uncertainty is the variability of \%SOC and BD across space---\emph{spatial heterogeneity}---which can be estimated by random sampling. When heterogeneity is high, more cores are required to estimate the total SOC stock in a given plot to within a given level of uncertainty. Larger samples are often needed to detect and quantify realistic stock changes \citep{kravchenko2011whole, stanley2023valid}. 
In some cases, stratification or balanced sampling can help to control for spatial heterogeneity economically \citep{gruijter2016farm, potash2023multi}. 
Compositing sampled cores before assay may also reduce costs, but introduces more opportunities for user error. Furthermore, the potential benefit is sensitive to the relative spatial heterogeneity and assay variability \citep{spertus2021optimal, stanley2023valid}. 
There can also be substantial \emph{temporal variability} in \%SOC in a given plot, which may confound random variation with a genuine trend in repeated measurements \citep{wuest2024temporal}. 
Seasonal variability can be controlled by fixing a sampling time (e.g., sampling in the same month or management period, year over year).

If samples are taken to a fixed depth, a decrease or increase in BD (e.g., from soil decompaction with increased organic matter content, or settling of soil particles following adoption of no-till) can result in an apparent decrease or increase in the C stock, even if it has not changed. 
The ESM procedure \citep{wendt2013equivalent}, attempts to circumvent this issue, but is not directly comparable to volumetric measurement with BD: SOC stock change is expressed on a mass-mass basis, with soil mass measurements used only to determine the depth window over which the average is taken. ESM introduces a modeling uncertainty, as the cumulative-soil-mass to SOC-mass relationship is estimated using a cubic spline. 
In rangelands and forests, large rocks and boulders can contribute uncertainty because they affect the volume of actual SOC-bearing soil. When rocks are not taken into account, stock and stock change estimates may be exaggerated. Unfortunately, precise accounting for rocks larger than an auger or BD ring is difficult and is not done in most studies. At the very least their presence should be noted. To avoid bias in volumetric SOC measurement, \%SOC and BD are best estimated on each core, multiplied together, and then averaged. If BD is measured via the ring method, \%SOC should also be measured on those rings in order to estimate the correlation between \%SOC and BD. Indeed, these quantities are typically negatively correlated with each other (e.g., \citet{curtis1964estimating}), so averaging before taking the product will yield stock estimates that are biased upwards. Upward bias may be common in reported estimates since joint measurement is laborious and rare.

\subsection{Study-level uncertainties}

Study-level uncertainties arise when causal conclusions are drawn.  
Studies that do not explicitly compare interventions cannot draw reliable causal conclusions. 
For instance, a study that measures SOC on a group of plots treated the same way is not suited to answer any causal question without making strong, untestable assumptions (for instance, that baseline SOC serves as a counterfactual for SOC at follow-up). 
In this paper, our focus is on field experiments---randomized controlled trials (RCTs)---but we briefly review uncertainties specific to observational studies as their shortcomings highlight the ideal role of experiments.

\subsubsection*{Observational studies}

Observational studies are subject to all of the uncertainties that arise in experiments plus additional sources of error due to \emph{confounding effects}, which occur when the receipt of treatment is influenced by baseline factors associated with the outcome. 
For example, if farmers target treatment to plots with higher baseline SOC (i.e., treatment selection is non-random), comparing observed treatment and control averages of SOC some time later will tend to make a treatment look more effective than it really is. 
When baseline SOC is measured, it is possible to adjust for its influence, and this extends to other possible confounders like weather, soil type, topography, and land use history. 
However, it is impossible to guarantee that all confounders are measured and properly taken into account, so conclusions may be highly sensitive to assumptions. 
Some observational, “space-for-time" studies do not capture data at baseline and followup, but measure two groups of plots at a single time point. 
These studies cannot be used to quantify SOC sequestration over time, and instead estimate the difference in treatment and control SOC storage at time of sampling. 
This type of study is especially sensitive to bias since, without baseline SOC, a crucial confounder is unobserved.

\subsubsection*{Randomized controlled trials}

Randomized controlled trials (RCTs) are experiments wherein interventions are randomly assigned to plots following a known probability distribution---the \emph{design}---possibly using additional control features such as blocking. This allows researchers to make unbiased estimates and inferences without strong modeling assumptions, and allows for rigorous causal inference.\footnote{ 
The key role of the randomization in causal inference, especially in field trials where complete control over confounders is impossible, has been recognized since RA Fisher and Jerzy Neyman’s pioneering work on the design of agricultural experiments in the early 20th century \citep{fisher1925statistical, neyman1923sur}.} 
We simplify our development by assuming the study is an RCT with $n$ equal-area plots enrolled and only two interventions: $n_1$ plots are assigned to treatment and $n_0 := n - n_1$ plots are assigned to control (the symbol “$:=$” means “is defined to be equal to,” as opposed to “happens to be equal to”). 
Our framework applies to studies with unequal plot sizes, unequal group sizes, or more than two treatments, but the notation is more complicated.

We ascribe two unknown \emph{potential outcomes} to each plot in the study: the stock it would have some years after receiving a treatment–if it receives the treatment–and the stock it would have after the same number of years if it does not receive the treatment. In symbols, let $Y_i(1)$ denote the SOC stock of plot $i$ at the end of the study if it receives treatment, and $Y_i(0)$ denote its stock if it receives control. If we knew both potential outcomes for each plot, we could compute any causal summary of interest, including the \emph{individual treatment effect} for plot $i$, $\tau_i := Y_i(1) - Y_i(0)$, the effect of treatment on the  amount of SOC stored in plot $i$. They are the target quantity for crediting schemes meant to incentivize management changes: plot $i$ should be credited in proportion to $\tau_i$.

In reality, we can only observe one potential outcome for each plot, depending on whether it was assigned to treatment or control. Hence, we cannot guarantee unbiased estimates of plot-level causal parameters. But the design allows us to make unbiased estimates of study-level causal parameters, like the \emph{average treatment effect} $\tau := \frac{1}{N} \sum_{i=1}^N \tau_i$, which summarizes the effect of a treatment across the whole study. 
For example, an average treatment effect of 5 Mg SOC ha$^{-1}$ means that on average,  5 Mg SOC ha$^{-1}$ additional SOC would have been stored during the study had every plot been treated. Scaling an estimate of that quantity by the number of hectares that were actually treated gives an estimate of the realized storage.

Even if there were no plot-level uncertainty, the average treatment effect would be unknown and estimates of it would be subject to noise arising from the random treatment assignment and \emph{inter-plot variability}. Inter-plot variability arises both from spatial heterogeneity in \%SOC and BD or ESM across plots, and from variable responses to treatment across plots. Studies need to enroll enough plots to control uncertainty from inter-plot variability, especially when attempting to detect relatively small treatment effects on a short time horizon \citep{polussa2026toward}. Additional design control measures like blocking or pairing plots based on possible confounding factors (e.g., location, soil type, historical land use, topography, etc) can help reduce the uncertainty from inter-plot variability without expanding the size and cost of a study, but need to be accounted for in data analysis.

\emph{Interference} arises when one plot’s treatment assignment affects another plot’s outcome (e.g., an amendment runs-off onto a down-slope control plot). \emph{Noncompliance} means that the treatment a plot received is not the treatment it was assigned (e.g., a plot assigned to AMP grazing is in fact grazed conventionally). Both will bias estimates of treatment effects and need to be controlled by the design protocol.

\subsection{Population-level (generalization) uncertainties}
\label{sec:population_uncertainty}

Population-level uncertainties arise when study results are extrapolated across space or time, which is essential to interpreting findings, designing new studies, and informing policy decisions. 
To generalize across time, researchers must assume that potential trajectories have a particular shape, for example, that SOC storage continues along a linear trajectory or reaches a plateau (equilibrium) and does not reverse. 
Linear trajectories are usually unrealistic, but are implicitly assumed when change estimates reported in Mg C ha$^{-1}$ y$^{-1}$ are simply multiplied by years. 
Furthermore, trajectories cannot depend on the absolute time at which a management change occurs: they must be \emph{stationary}. 
For example, a stationary treatment effect implies that switching the population from control to treatment now would create the same trajectory as making the same switch in 20 years. 
This assumption is reasonable in the absence of major changes in a population over time, but such changes could be caused by climate change, for instance. 
There is no way to account for non-stationarity without making other untestable assumptions, but results can be checked for their sensitivity to the stationarity assumption.

On the other hand, generalizing across space---also called \emph{upscaling}---can be made more rigorous by design. 
To ensure external validity, the plots enrolled in a study should be representative of the larger population to which the researchers seek to extrapolate the findings. 
The ideal strategy is to randomly sample plots from the population of interest and enroll them into an RCT. 
Such random enrollment is rarely feasible in practice, and studies more often enroll plots by systematic or convenience samples. 
When enrollment is done in this way, generalization must be based on informal reasoning. 
If auxiliary data such as climate, topography, soil type, or land use history are available, formal methods can be applied to adjust the study to be more representative of the population. 
For example, if a particular soil type is under-represented in a study compared to a population of interest (and soil type is thought to moderate the treatment effect), then plots with that soil type in the study should receive a higher weight when estimating the average treatment effect across the whole population \citep{egami2023elements}.

\section{Statistical model}
\label{sec:causal_model}
 
Here we outline an approach for defining and estimating SOC storage potential in an experimental context. As above, plot $i$ is some volume of soil that can be treated or not. There are $N$ plots in the population of interest, and $n$ plots in the study. Time is denoted by $t \geq 0$, where $t=0$ is baseline; we will generally use years as the unit of time. To simplify our exposition, we assume a binary treatment with only two levels, but our framework is readily generalized to treatment with multiple levels or to continuous treatments. Each plot $i$ in the population has two \textit{potential trajectories} of SOC stock denoted $y_i(1,t)$ and $y_i(0,t)$. We define \textit{baseline time} to be $t=0$ and \textit{baseline stock} to be $y_i(0,0) = y_i(1,0)$. We will express time in years, so $y_i(1,5)$ is the SOC stock in plot $i$ 5 years from baseline if that plot receives treatment. A \textit{potential outcome} (PO) $y_i(z)$ is derived from a potential trajectory by fixing\footnote{It could also be defined as a (weighted) average over a time window, a maximum, or another functional that marginalizes over time. Such definitions smooth out seasonal temporal variability but are harder to measure, requiring repeated sampling over time.} time and dropping it from the notation. For example, we might define $y_i(0) := y_i(0,5)$ for all $i$ to be the stock after 5 years if the plot is assigned to the control group. Each plot in the population also has a length-$p$ vector of covariates observed at baseline and denoted $\bs{x}_i$. The covariates represent possible \textit{moderators} of SOC storage potential, including baseline SOC, soil type, clay content, land-use history, or NDVI.

\subsection{Target Quantities}
 
Let $\bs{z} := [z_1,\ldots,z_N]$ be a \textit{treatment policy}, a length-$N$ vector whose $i$th component is 1 if plot $i$ is assigned to treatment and 0 if not. Suppose the volume of the plots is equal (if volumes are unequal, parameters and estimates must be weighted by them). Under a given treatment policy $\bs{z}$, the (population) \textit{average potential outcome} is $\bar{y}(\bs{z}) := \frac{1}{N} \sum_{i=1}^N y_i(z_i)$, representing the average mass of SOC in the population some years after treatment policy $\bs{z}$ has been implemented. The total SOC stored is just $N \times \bar{y}(\bs{z})$.
 
The canonical goal of causal inference is to estimate contrasts between treatments. For instance, the (population) \textit{average treatment effect} is:
$$\tau := \bar{y}(\bs{1}) - \bar{y}(\bs{0}) = \frac{1}{N} \sum_{i=1}^N \left [ y_i(1) - y_i(0) \right ] = \frac{1}{N} \sum_{i=1}^N \tau_i$$
where $\tau_i = y_i(1) - y_i(0)$ is the individual treatment effect for plot $i$. This is the amount of additional SOC storage on average per plot in the population if every plot were treated compared to the average SOC storage if no plot had been treated ($N \times \tau$ is the total additional SOC stored). If the average treatment effect is positive, then as a blanket policy, the treatment stores more SOC than the control. But if there are financial or other constraints on resources (so that not all $N$ plots can be treated) or if some plots lose SOC under treatment, such a blanket policy might not be best: a tailored policy that targets plots expected to respond the best to treatment might yield greater return per investment.
 
We can now define the optimal policy in terms of the average potential outcomes. Suppose policy vector $\bs{z}$ incurs overall cost $c(\bs{z})$ and the total budget is $C_0$. The \textit{treatment portfolio}:
$$\mathcal{P} := \{\bs{z} : z_i \in \{0,1\}, c(\bs{z}) \leq C_0 \}$$
is the set of all treatment policies that meet the budget constraint. A policy $\bs{z}$ is called \textit{feasible} if $\bs{z} \in \mathcal{P}$. A portfolio is \textit{unrestricted} if $\mathcal{P} = \{0,1\}^N$, so that any policy is feasible. If each plot has cost $c_i(z_i)$ for action $z_i$ and costs are additive across plots, then we can decompose $c(\bs{z}) = \sum_{i=1}^N c_i(z_i)$. We return to this assumption in the discussion, but note that additivity of costs across plots may not be realistic if there are economies of scale or savings from treating proximate plots.
 
The \textit{optimal average potential outcome} is $\bar{y}^* = \max_{\bs{z} \in \mathcal{P}} \bar{y}(\bs{z})$. An \textit{optimal policy} $\bs{z}^*$ is any treatment policy $\bs{z}^* \in \mathcal{P}$ that achieves the optimal average potential outcome, i.e., $\bar{y}^* = \bar{y}(\bs{z}^*)$. The optimal policy need not be unique. For instance, if the treatment and control have equal costs and equal effects for every plot, then any feasible policy is optimal. If $\bar{y}^*$ is much larger than both $\bar{y}(\bs{1})$ and $\bar{y}(\bs{0})$, then the average treatment effect is not the most relevant parameter for treatment decisions: we could achieve a substantially larger return by targeting treatments to plots. It is also possible that either of the \textit{uniform policies} $\bs{z} = \bs{1}$ or $\bs{z} = \bs{0}$ are infeasible, in which case the policy-maker is forced to triage the interventions.

\section{Design, data, and estimation}
\label{sec:data_estimation_inference}
 
We assume $n$ plots are enrolled in the study by simple random sampling without replacement\footnote{
This is not typical of real-world agricultural experiments, which most often enroll plots deterministically. In practice, features that influence enrollment should be recorded (a) to estimate `as-if' enrollment probabilities and use them to adjust the analysis by artificially up-weighting plots that are under-represented in the population or (b) to form conditional average treatment effect estimates for the entire population and impute the missing potential outcomes \citep{egami2023elements}.
} from a larger population of plots. Let $\mathcal{S} = \{S_1,\ldots,S_n\}$ be the collection of $n$ random indices of the enrolled plots. For each plot $i \in \mathcal{S}$, there is a covariate vector $\bs{X}_i$, a potential outcome $Y_i(0)$ that will be observed if the plot receives control, and a potential outcome $Y_i(1)$ that will be observed if the plot receives the treatment. Let $Z_i = 1$ if study plot $i$ is assigned to treatment and $Z_i = 0$ if it is assigned to control. Plots are assigned to treatment with a known joint probability distribution. In a completely randomized experiment, the treatment group contains exactly $n_1$ plots and the control group contains $n_0$, where $n_1 + n_0 = n$; treatment is assigned in such a way that every subset of $n_1$ of the plots is equally likely to be the treated group. The (marginal) chance plot $i$ is assigned to treatment is $\mathbb{P}(Z_i = 1) = n_1/N$ and the chance it is assigned to control is $\mathbb{P}(Z_i = 0) = n_0/N$. After the study, we record the \textit{observed outcome} $Y_i := Y_i(Z_i)$ for each plot in the study. All in all, the observed dataset is $\{(Y_i, Z_i, \bs{X}_i)\}_{i=1}^n$ recording the observed outcome, treatment assignment, and covariates for each plot.
 
If plots are selected at random from the population and then assigned at random to treatment, the average treatment effect can be written $\tau = \mathbb{E}[Y_i(1)] - \mathbb{E}[Y_i(0)]$. We cannot estimate the individual treatment effect for a plot, but we can estimate a \textit{conditional average treatment effect} (CATE), $\tau(\bs{x}) := \mathbb{E}[Y_i(1) \mid \bs{x}] - \mathbb{E}[Y_i(0) \mid \bs{x}]$, the amount by which treatment is expected\footnote{Formally, the conditional average treatment effect is $\tau(\bs{x}) := \mathbb{E}[Y_i(1) - Y_i(0) \mid \bs{X}_i = \bs{x}]$, the expected value of the difference in potential outcomes conditional on the covariate value $\bs{x}$. The expectation is with respect both to random sampling from the population and to random treatment assignment. See \citet{kunzel2019metalearners}.} to increase SOC storage in a plot whose covariate values are $\bs{x}$.
The best estimator of $\tau(\bs{x})$ that minimizes mean squared error is the best estimator of the individual treatment effects \citep{kunzel2019metalearners}, which in turn provide better estimates of the optimal policy \citep{qian2011performance}.
 
CATEs are estimated by $\hat{\tau}(\bs{x}) := \hat{\mu}_1(\bs{x}) - \hat{\mu}_0(\bs{x})$, where $\hat{\mu}_1(\bs{x})$ is a regression of follow-up SOC $Y_i$ on covariates $\bs{X}_i$ for the $n_1$ treated units, and $\hat{\mu}_0(\bs{x})$ is the same regression among the $n_0$ control units. See \Cref{fig:appendix_cate_intuition} for an illustration.
The regressions can be fit using ordinary least squares (OLS) \citep{ding2019decomposing} or machine learning algorithms such as random forests or $K$-nearest-neighbors \citep{kunzel2019metalearners, athey2021policy}, both of which have consistent standard error estimates and pointwise confidence bands for uncertainty quantification. Partially pooled methods---including mixed-effects and Bayesian hierarchical models---could also be used, but are less well studied and may require additional model-based assumptions. Machine learning may be especially useful when there are many covariates, the experiment is large, and there could be high-order interactions between covariates. OLS may be more suitable when the form of the moderating effects is well structured by theory, and when interpretable linear estimates are important. In particular, baseline moderation can be evaluated by using OLS with baseline \%SOC as the covariate; if baseline moderation is suspected to be nonlinear, a polynomial or a spline basis expansion can be created and fit by OLS. The estimated CATE $\hat{\tau}(\bs{x})$ can be plotted over a range of baseline \%SOC or other features.
 
Given estimates of the CATE, we can estimate any given population average potential outcome by
$$\hat{\bar{y}}(\bs{z}) := \frac{1}{N} \sum_{i=1}^N \left [ \hat{\mu}_1(\bs{x}_i) z_i + \hat{\mu}_0(\bs{x}_i) (1-z_i) \right ].$$
That is, we impute every potential outcome in the population using the regressions, choose the potential outcome for each unit that corresponds to its management under policy $\bs{z}$, and average them. The optimal policy is estimated by ``plugging-in" the population average potential outcome estimates:
$$\hat{\bs{z}}^* \in \argmax_{\bs{z} \in \mathcal{P}} \hat{\bar{y}}(\bs{z}).$$
If the portfolio is unrestricted (if there are no cost or logistical constraints), the optimal policy simply treats every plot where $\hat{\tau}(\bs{x}) > 0$. If the portfolio can treat at most $K$ plots in the population, an approximately optimal policy treats the $K$ plots with the highest estimated CATE (or fewer than $K$, if there are not $K$ with positive estimated CATE). For instance, if there is negative baseline moderation in the study, then an optimal policy would allocate treatment to the plots with the $K$ lowest baseline SOC measurements in the population. In general, there are $2^N$ policies with a binary treatment, but the optimal policy can be found for large $N$ by dynamic programming (see discussion). It is also possible to estimate an optimal policy using treatment assignment and enrollment probabilities rather than CATE estimates \citep{kitagawa2018who} or to combine the two \citep{athey2021policy}, but CATE estimates are needed to evaluate moderation.
The bootstrap can be used to construct a confidence set for the estimated value of the estimated optimal policy (not the estimated value of the \emph{true} optimal policy, which is not identified; see discussion).
 
\paragraph{A note on measurement error}

In the context of a larger experiment, both core and plot-level uncertainties can be considered \textit{measurement error} that affects both observations of $Y_i$ and components of $\bs{X}_i$ (e.g., baseline \%SOC). 
High measurement error will bias moderator estimates towards 0, so controlling measurement error through intensive plot-level sampling is especially important to obtaining refined estimates of storage potential.

\section{Empirical studies}
\label{sec:simulations}
 
\subsection{California compost experiment}
 
The data were collected as part of an experiment started in 2016 \citep{silver2018carbon}. The study enrolled pairs of 30 m $\times$ 62.5 m plots at 7 different rangeland sites (14 plots total) across California, representing a wide range of grassland soil orders typical to the region (Alfisols and Mollisols). Within each pair, one plot was chosen at random to be treated with a thin layer of compost, while the other plot served as a business-as-usual control. Plots were measured for \%C by drawing 5 randomly located soil cores stratified along a diagonal transect to a depth of 10 cm. Cores were assayed for \%C by elemental analysis; inorganic carbonates were minimal so that \%C $\cong$ \%SOC. This entire measurement procedure was carried out in 2016 just before compost was applied, and repeated every subsequent year until 2019. We dropped one site that was not resampled in 2019. There were thus 60 cores overall, 30 with the compost amendment applied. From a design-based perspective, analyzing core-level data is pseudo-replication because plots were the primary unit randomized \citep{hurlbert1984pseudoreplication}. Thus, we also ran this analysis at the plot-level.

We implemented SOC storage policy optimization with the California data, taking follow-up \%SOC from core-level 2019 measurements and baseline \%SOC from core-level 2016 measurements. We estimated the CATE on follow-up SOC moderated by baseline SOC using linear regression (OLS). We estimated 90\% pointwise confidence bands for the CATE using the normal approximation with heteroskedasticity robust standard error estimates \citep{ding2019decomposing}. We plotted the estimated CATE function alongside a “simple regression" estimate of baseline moderation, computed by regressing the change in SOC (follow-up SOC minus baseline SOC) on baseline SOC for all cores \citep{slessarev2023initial}. We then computed the total returns on three policies: the unconstrained optimal policy, which treats every unit with a positive CATE; a treat-10\% policy, which is budget-constrained to assign at most 10\% of units to treatment with the rest on control; and a blanket policy, which either assigns all units to treatment or assigns all units to control, depending on which has a larger average treatment effect estimate. We quantified uncertainty in the estimated values of the policy estimates using the stratified bootstrap, resampling (with replacement) an equal number of cores from each treatment arm, re-estimating the optimal policy and its value in each of 1000 bootstrap replicates. A 90\% confidence interval was estimated as the 5th and 95th percentiles of those replicates.

\begin{figure}
    \centering
    \includegraphics[width = \textwidth]{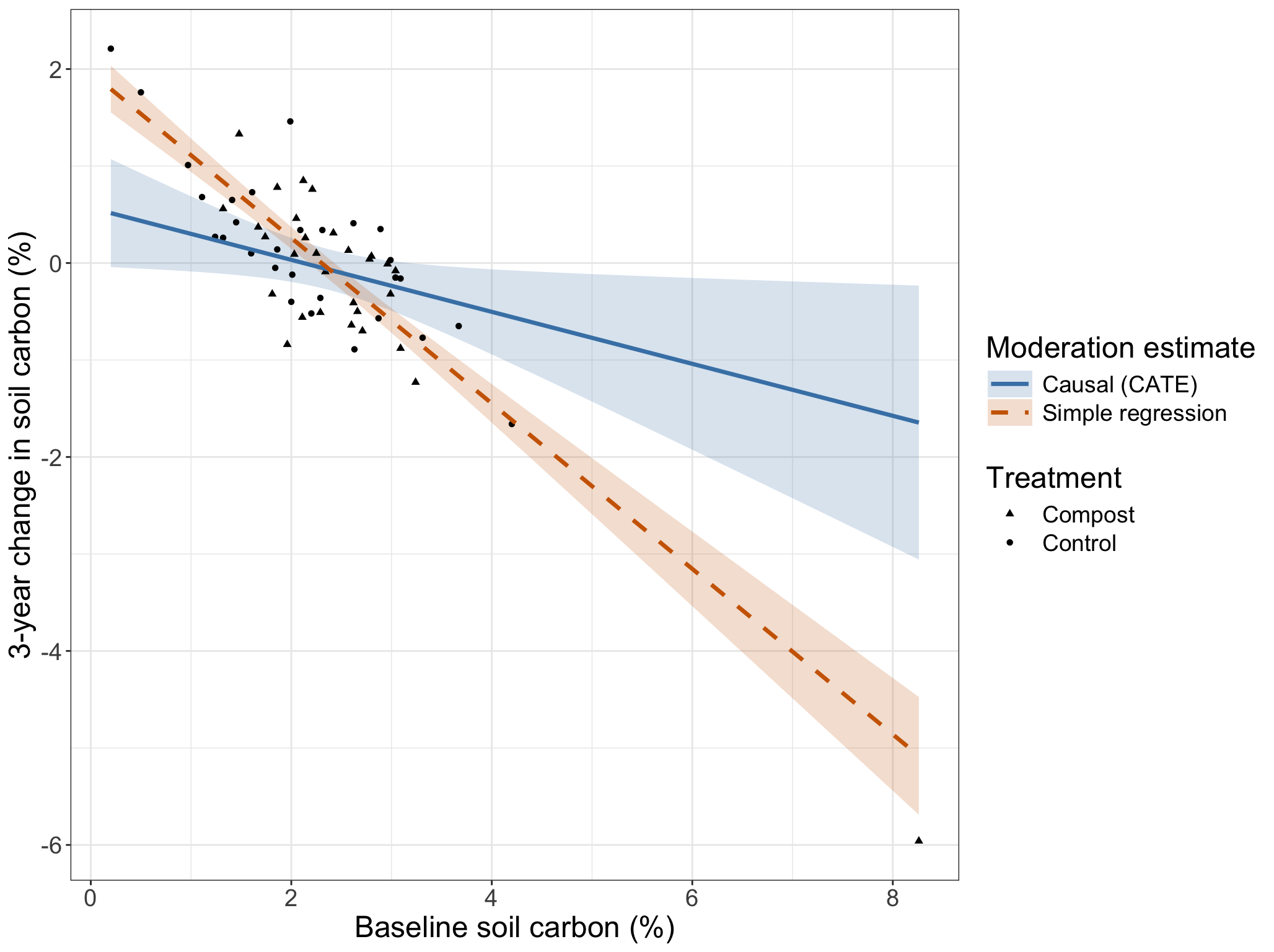}
    \caption[Baseline moderation on California rangeland]{
    Observed soil core percent carbon measurements and C saturation estimates from a compost experiment on California rangeland. At the start of the experiment, baseline C concentrations (\%) were measured (X-axis) and points were assigned to treatment-free control (circles) or to receive a compost amendment (triangles) following a randomized paired-plot design. SOC concentrations were measured again after 3 years, and the change from baseline was computed (Y-axis). A `simple regression' of C change on baseline C for every core indicates a speciously strong baseline moderation effect (orange line). Causally estimating the conditional average treatment effect (CATE; blue line) indicates a smaller moderating role for baseline SOC. The ribbons are 90\% confidence bands. The CATE estimate indicates that a policy applying compost only to plots with low ($\leq 2\%$) baseline C will store more carbon than either of the ``blanket policies'' that apply compost to all plots or to no plots. Removing the exceptional data point at 8\% baseline C causes the CATE estimate to become flat near 0, but does not change the simple regression estimate (\Cref{fig:appendix_saturation_outlierremoved}).
    }
    \label{fig:nrcs_saturation}
\end{figure}

\Cref{fig:nrcs_saturation} plots observed data and the CATE estimates and confidence bands. The CATE estimate decreases as baseline SOC increases, suggesting mild baseline moderation. 
However, the moderation is considerably lower than the simple regression estimate, which (speciously) indicates a very strong moderating effect of baseline C \citep{slessarev2023initial}. 
Based on this data, targeting treatment to plots with low ($\lesssim 2\%$) baseline C would yield slightly more storage than uniformly assigning plots to either treatment or control.
The unconstrained optimal policy treats 45\% of units with compost and achieves a follow-up average \%SOC of 2.46 [95\% CI: 2.32, 2.57], which is 103.7\% of baseline average SOC. 
The best “treat 10\%" policy treats the 6 cores with the lowest baseline SOC, and achieves a follow-up average \%SOC of 2.40 [95\% CI: 2.26, 2.53], which is 102.4\% of baseline average SOC. 
The best uniform policy dictates that all units receive compost—the arm with the largest average treatment effect in the study—and achieves a follow-up average of 2.36\% [95\% CI: 2.27, 2.51], which is 101.0\% of baseline. 
The plot-level analysis (\Cref{fig:appendix_saturation_plotlevel}) also indicated baseline moderation---with the CATE estimate lower than the simple regression estimate---and higher gains over time for all policy estimates (106\% relative to baseline for the unconstrained policy; 105\% for the constrained policy, and 104\% under control).

The CATE estimates reveal a moderating effect on baseline, while the optimal policy estimates and confidence intervals suggest only marginal gains to optimizing the policy. 
That is, given the choice of assigning all plots to treatment, all plots to control, or according to an estimated optimal policy, a decision maker would be essentially indifferent without more information (e.g., a larger or longer experiment, or a stronger cost model). 
Moreover, these results were subject to different climate conditions across sites and to different weather events over the study period. 
In reality, estimating the weather or climate impacts of organic amendments rigorously is complex and requires more information than SOC stock measurements. 
Process-based models are typically required to consider the alternative fates of amendment C, quantify non-CO$_2$ greenhouse gas fluxes, and partition direct and indirect effects of amendments \citep{delonge2013lifecycle, mayer2022climate}. 
Our analysis here does not take these factors into account, but nonetheless serves as a proof of concept that CATE estimates can be used to inform policy optimization.

\subsection{Simulations}
 
We extended our evaluation of these optimal policy estimators in a simulation study. Namely, we imagined a hypothetical field experiment with $N = \{20, 40, 100, 200\}$ plots, with a binary treatment assigned at random to each: 50 plots received treatment and 50 were on control. After fitting models to learn the CATE functions in the study, we estimated storage potential and optimal policies in a separate population of 500 plots. The study was assumed representative of the population, so they were drawn from the same data generating process.
 
We drew baseline \%SOC from a gaussian with mean 3 and standard deviation 1, truncating the values below 0. Control potential \%SOC at follow-up was set equal to baseline \%SOC. Treatment potential \%SOC at follow-up was determined by an ATE $\tau$ and one of three moderator scenarios: no moderation, baseline moderation, and complex moderation.
\begin{itemize}
    \item Under \textbf{no moderation}, treatment \%SOC was equal to control \%SOC plus a constant treatment effect $\tau_i = \tau \in \{0\%, 0.5\%, 1\%, 2\%\}$, as an absolute (that is, not relative to baseline) shift in \%SOC compared to control.
    \item Under \textbf{baseline moderation}, the ATE was as above, but we added a linear effect of baseline \%SOC on each plot. We first normalized baseline to $[-1,1]$. For normalized baseline $B_i$, we modeled: $y_i(1) = y_i(0) + \tau + \beta \times B_i$, where $\beta \in \{-0.1, -0.5\}$ indicated a small or a large moderating effect: higher baseline implies lower treatment effect. We refer to $\beta = -0.1$ as ``weak baseline moderation'' and $\beta = -0.5$ as ``strong baseline moderation.''
    \item Under \textbf{complex moderation}, we added 3 moderators: baseline \%SOC, a balanced binary indicator proxying soil type, and a uniform $[-1,1]$ random variable proxying NDVI. We modeled $y_i(1) = y_i(0) + \tau - B_i \times S_i - B_i \times C_i$. Treatment thus had less effect on plots with soil type 1 and high baseline \%SOC, and on plots with high NDVI and high baseline \%SOC.
\end{itemize}
We estimated the average SOC storage of the best blanket policy, the linear ordinary least squares (OLS) optimal policy estimate, and a random forest (RF) machine learning optimal policy estimate. Our RF estimate used the causal forest estimator implemented in the \texttt{grf} R package (version 2.6.1), which fits separate trees using treatment assignment as a splitting criterion. We also evaluated the returns of the OLS and RF ``treat-best-10\%'' policies, which only treats the plots with the highest SOC storage potential and leaves the rest as controls. 
As a benchmark, we computed the return of an \textit{oracle policy}, which takes the true largest potential \%SOC for each plot and sums them across plots. 
We ran 100 simulations of every scenario and recorded the average \%SOC return under each policy. We calculated the ratio of the returns of various optimal policy estimates over the returns of the best blanket policy---one that assigns all plots to treatment or all plots to control.
 
\begin{figure}
    \centering
    \includegraphics[width=\linewidth]{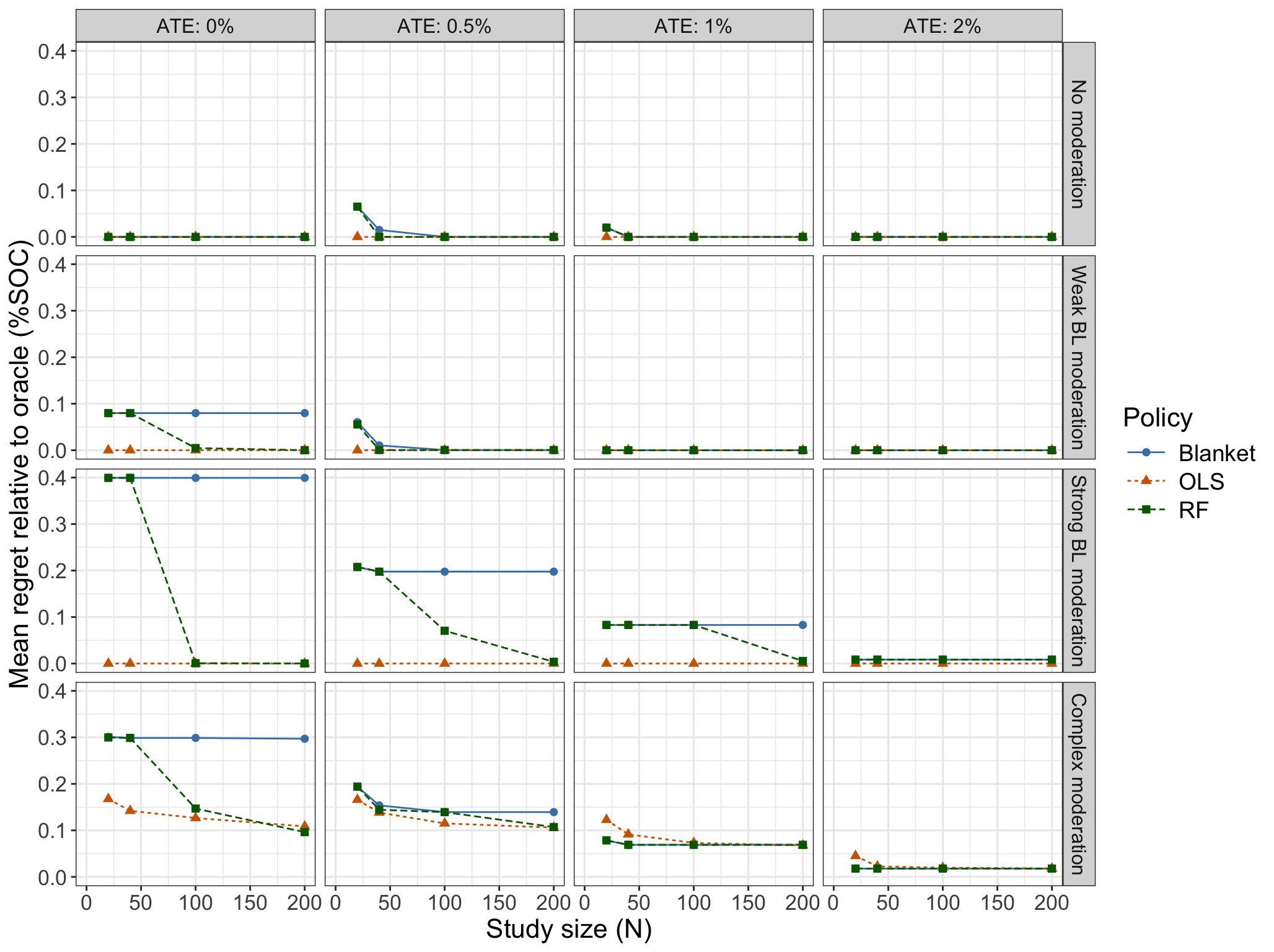}
    \caption{Regret (y-axis) of three optimal policy estimators in terms of their expected gap (regret) in average \%SOC across the population compared to the oracle policy, plotted over various study sizes (x-axis). 
    A policy with lower regret is better, and 0 is optimal. Regret was estimated as the average gap from the oracle policy over 500 simulations. 
    Different moderation scenarios, described in the main text, appear in the panel rows; different average treatment effects (ATEs) are in the panel columns. BL = baseline; OLS = ordinary least squares; RF = random forest.
    }
    \label{fig:regret_plot}
\end{figure}

\Cref{tab:percent_improvement} displays the return ratios as the percent of the blanket policy return in various scenarios. The OLS-estimated optimal policy typically matches the return of the oracle policy and beats the RF-estimated optimal policy, especially in small studies. \Cref{fig:regret_plot} displays the \textit{regret} of these policies (the gap between their return and the oracle return) as a function of $N$, over different moderation scenarios and average treatment effect sizes. OLS and RF improve on the blanket policy when there is moderation, but RF needs substantially larger studies to realize the gains of OLS. Large ATEs ``wash out" the advantages of the more sophisticated estimators: since all individual treatment effects are positive, the blanket policy is optimal.

\begin{table}[!ht]
\centering
\centering
\begin{tabular}[t]{lrllllll}
\toprule
ATE & N & Scenario & Oracle & OLS & RF & OLS-10\% & RF-10\%\\
\midrule
0.0\% & 20 & No moderation & 100.0\% &  100.0\% & 100.0\% & 100.0\% & 100.0\%\\
 &  & Weak baseline moderation & 102.7\% & \cellcolor{lime}102.7\% & 100.0\% & 101.2\% & 100.0\%\\
 &  & Strong baseline moderation & 113.3\% & \cellcolor{lime}113.3\% & 100.0\% & 105.8\% & 100.0\%\\
 &  & Complex moderation & 110.0\% & \cellcolor{lime}104.4\% & 100.0\% & 102.0\% & 99.9\%\\
\cmidrule(lr){2-8}
 & 200 & No moderation & 100.0\% & 100.0\% & 100.0\% & 100.0\% & 100.0\%\\
 &  & Weak baseline moderation & 102.7\% & \cellcolor{lime}102.7\% & \cellcolor{lime}102.7\% & 101.2\% & 101.1\%\\
 &  & Strong baseline moderation & 113.3\% & \cellcolor{lime}113.3\% & \cellcolor{lime}113.3\% & 105.8\% & 105.8\%\\
 &  & Complex moderation & 109.9\% & 106.3\% & \cellcolor{lime}106.7\% & 102.6\% & 104.8\%\\
\midrule
1.0\% & 20 & No moderation & 100.5\% & 100.5\% & 100.0\% & 77.9\% & 77.8\%\\
 &  & Weak baseline moderation & 100.0\% & 100.0\% & 100.0\% & 78.4\% & 77.5\%\\
 &  & Strong baseline moderation & 102.1\% & \cellcolor{lime}102.1\% & 100.0\% & 81.9\% & 77.5\%\\
 &  & Complex moderation & 102.0\% & 98.9\% & 100.0\% & 79.1\% & 77.5\%\\
\cmidrule(lr){2-8}
 & 200 & No moderation & 100.0\% & 100.0\% & 100.0\% & 77.5\% & 77.5\%\\
 &  & Weak baseline moderation & 100.0\% & 100.0\% & 100.0\% & 78.4\% & 78.4\%\\
 &  & Strong baseline moderation & 102.1\% & \cellcolor{lime}102.1\% & 101.9\% & 81.9\% & 81.8\%\\
 &  & Complex moderation & 101.7\% & 100.0\% & 100.0\% & 79.5\% & 81.1\%\\
\bottomrule
\end{tabular}
\caption{
Geometric mean (over scenarios and study sizes) of the ratio of optimal policy returns over the returns of the best blanket policy (``treat every plot").
Four methods were used to estimate storage potentials and choose a policy: ordinary least squares (OLS), random forest (RF), OLS treating at most 10\% of plots (OLS-10\%), and RF treating at most 10\% of plots (RF-10\%).
The oracle knows all potential outcomes and represents the true optimal policy.
The other columns represent various estimators of storage potential.
In each row, the top method(s) are highlighted in green, if they exceed the average return of the blanket policy by more than 1\%.
ATE = average treatment effect as additional percent soil organic carbon (\%SOC).}
\label{tab:percent_improvement}
\end{table}

\section{Discussion}
\label{sec:discussion}
 
We provided an integrated review of uncertainties in studies targeting SOC storage along with a design-based causal model for estimating treatment effects and optimizing storage across a population of interest. We proposed methods to estimate SOC storage potential and then approximate an optimal policy using those estimates. In simulated field experiments, ordinary least squares (OLS) bested the standard approach of uniformly applying the management with the largest average treatment effect as a blanket policy. It also performed better than a more sophisticated machine learning method based on random forests, except when the studies were infeasibly large. Indeed, existing field experiments are typically fairly small: to our knowledge none has enrolled 200 plots, the upper limit in our simulations. Recent efforts to compile and harmonize data across experiments may approximate much larger designs and yield new insights about SOC storage potential, sequestration, and optimal management \citep{bybeefinley2024deriving}.

We recommend that SOC researchers use our framework with the OLS estimator to evaluate moderation and potential gains to policy optimization, as a complement to reporting treatment effects in terms of average sequestration rates. As with all statistical approaches, these methods will add the most insight when signals are large–supported by identifying moderators through scientific reasoning and running long term studies that allow effects to manifest–and when noise is small–supported by careful power analysis in light of the sources of uncertainty discussed above. There are various limitations of our work, and exciting possibilities to extend it.

\subsection{Other study designs and potential pitfalls}
 
We assumed a specific design: a randomized controlled trial (RCT) with units sampled uniformly at random from a broader population of interest. This represents an ideal wherein internal and external validity are justified by the design alone. Sometimes such a design is infeasible in real-world agricultural experiments, but some of the assumptions can be relaxed.
 
The assumption of uniform sampling into a completely randomized experiment is not necessary: there exist estimators that retain unbiasedness and inferential validity under a much broader range of designs, including non-uniform sampling and assignment, stratification or blocking, rerandomization, and cluster sampling or treatment assignment \citep{imbens2015causal, egami2023elements}. Furthermore, the plots in the experiment may constitute a convenience sample from the larger population or may not be embedded in a population at all. In that case all inferences may be confined to the experiment itself. We emphasize the importance of careful consideration of the population to which results are being extrapolated, and of replicating studies. Genuine replication is the strongest way to establish that causal effects generalize across contexts.
 
Purely observational studies---including longitudinal and space-for-time designs---are more problematic, requiring hypothetical populations and sampling premised on unverifiable assumptions. The design-based view of observational causal inference aims to reconstruct an RCT by weighting or matching on assumed assignment probabilities, then applies simple analyses in an effort to reveal weaknesses in the study \citep{rosenbaum2002observational}.

\subsection{A causal view on baseline moderation}
 
The method we developed can be used to evaluate whether baseline SOC levels moderate SOC accrual. A large body of past work has failed to provide robust evidence of baseline moderation, in large part because baseline moderation has not been situated within a causal model, leading to confusion \citep{slessarev2023initial}. In short, investigators have taken baseline SOC $B_i$ and followup SOC $Y_i$, computed the difference $D_i := Y_i - B_i$, regressed $D_i$ on $B_i$, and interpreted a negative regression coefficient as evidence that baseline SOC is negatively associated with change. This is not a meaningful result: the difference between any two independent random variables will be correlated with either one of the original random variables. Our approach instead formalizes the baseline moderation hypothesis in a causal model. When there is negative baseline moderation, a digital soil map of baseline SOC could be employed to locally target interventions to low SOC areas in order to maximize SOC storage at a fine scale (e.g., across a single farm).

\subsection{Cost models}
 
In characterizing the available treatment portfolio $\mathcal{P}$, we assumed that the overall cost of a policy was a linear and additive function of the costs of each individual treatment. This assumption creates a linear budget constraint set and makes the optimization problem (a `0-1 knapsack problem') solvable exactly by dynamic programming or approximately by linear programming with a relaxation that allows fractional assignments.
 
In many cases, a linear cost model is not appropriate. For example, policies can exhibit economies of scale so that marginal costs diminish as more plots receive a given treatment. All else being equal, diminishing marginal costs tend to suggest a more parsimonious policy, where only a few intervention levels are prescribed in relatively equal proportions. In the extreme, it suggests treating all plots the same, i.e. reverting to a blanket policy. The true costs of interventions will often depend on geography, especially for treatments involving shipping costs (e.g., organic amendments). Intuitively, it is cheaper to spread $x$ amount of an amendment on the same plot then to distribute $x/2$ to two plots. How much cheaper depends on the distance between them and the distance from the source of the amendment.
 
These considerations could make the optimization much more complex and potentially intractable, even if every potential outcome were known. Brute-force solutions are generally impossible, since, for a treatment with $K$ levels, they involve enumerating $K^N$ outcomes. If geography is a primary concern, the population could be partitioned into $G$ nearby clusters with all plots in a cluster constrained to receive the same treatment, reducing the burden of enumeration to $K^G$. Defining an accurate cost model and finding tractable routines for optimization is an important area for further research.

\subsection{Policy inference}
\label{sec:inference}
We summarized uncertainty through a bootstrap confidence interval on the value of the best blanket policy and on the \emph{estimated} optimal policy. Design-based inference on the true optimal policy is impossible: model-based assumptions (e.g., that individual treatment effects are linear transformations of observed covariates plus mean-zero noise) are needed. Inference on the estimated optimum produces a confidence set containing all policies close to the `sample best', so that we are essentially indifferent to the policy choice within that set. 
The policy maker could then pick the least costly policy from those.
Alternatives to bootstrapping may include multiple comparisons with the sample best \citep{hsu1996multiple}, inference on winners \citep{andrews2024inference}, or policy regret bounds \citep{athey2021policy, kitagawa2018who}.

\subsection{Multivariate outcomes}
 
It may be of interest to include additional potential outcomes in management policy design, like crop yield or additional soil health indicators. To approximate a single optimal policy, outcomes need to be commensurated---for instance by converting both yields and SOC storage to dollars using commodity and C prices. Alternatively, an optimal policy could be approximated for each outcome separately, and those policies compared. If there is overlap in the confidence sets for two or more estimates, an element of their intersection could be chosen to approximately optimize all outcomes simultaneously.

\subsection{Uncertainties at the policy level}
 
In our review of measurement approaches in Section 2, we did not address uncertainties that arise specifically around policy design, including additionality, permanence, and leakage. \textit{Additionality} is often defined to mean that the policy creates SOC sequestration that would not have occurred otherwise, but this generally assumes the effectiveness of interventions. Separating the implementation of an action from its effectiveness, we define \textit{policy additionality} to mean that the policy stimulates an action that would not have occurred otherwise. For example, when a land manager is paid to stop tilling, they do, and they would not stop if they were not paid: the policy causes the management change to occur. In this respect, additionality is closely related to compliance in experiments, and may be evaluated by considering existing regulatory, financial, physical, and social (dis)incentives to adopt an intervention and by recording information about compliance rates. While the scientific effects of an intervention are best estimated when there is complete compliance, policy additionality is best estimated when the study reflects the level of compliance that would occur if the policy were actually implemented.
 
\textit{Permanence} (or durability) often refers to the longevity of additional stored SOC \citep{oldfield2022crediting, smith2005overview, thamo2016challenges}, where a return to baseline SOC before some time horizon is called a reversal. Often 100 years is used as a time horizon \citep{oldfield2022crediting}, although the fate of stored carbon has the potential to affect global climate for longer \citep{mayer2026cost}. We differentiate between (a) the permanence of an intervention and (b) the permanence of stored SOC. Concept (a) is an additional population-level policy concern and uncertainty, which, like additionality, can be studied in the present by evaluating the rate of management reversals in a longitudinal study. Forecasting permanence of interventions in the future is far more challenging and carries systemic risks (e.g., related to lapses in institutional support for climate commitments or major changes in the agricultural economy \citep{mayer2026cost}). Concept (b) is an equally complex question that needs to be addressed with additional scientific knowledge about the SOC trajectory after an intervention is implemented and sustained. Long-term studies are particularly useful for furnishing such knowledge, but must be carefully generalized to the target population.
 
Finally, \textit{leakage} refers to the externalities of a particular intervention in terms of its greenhouse effect. An intervention that stores SOC but releases a large amount of methane exhibits leakage (e.g., in flooding fields for conversion to rice cultivation \citep{minami1994methane, nitta2022rice}), as does an intervention that burns fossil fuel as part of a supply chain (e.g., in transporting compost long-distances \citep{delonge2013lifecycle, silver2018carbon}), or an intervention that drives C-emitting land use change elsewhere (e.g., displacing food production with bioenergy crops \citep{plevin2010greenhouse}). It is critical to account for leakage through life-cycle assessment, tracking on-farm fossil fuel use, and measuring externalities like methane, nitrous oxide, and nitrogen oxide production. Again, comparison to appropriate counterfactual scenarios is crucial and can be formalized using potential outcomes.

\subsection{Conclusions}

As discussed in our introduction, the ``optimality'' of a sequestration-optimal policy really depends on values and context.
Even taking that goal for granted and ignoring uncertainty, conditional average treatment effects can produce policies that are far from optimal because sequestration potentials and individual treatment effects may depend on features that are unmeasured (e.g., moisture content, minerality, microbial activity, land-use history, etc). 
In that case, a local decision-maker (e.g., an experienced farmer or scientist) could make better judgments. 
Algorithms may support, but should not replace, situated knowledge.

\section*{Statements and Declarations}

\subsection*{Code and Data}
 
All code used to implement our data analyses and simulations are available on Github. 
Our code includes a modular function structure to flexibly implement optimal policy estimation. 
See \url{https://github.com/spertus/SOC-storage-potential}. 
The Github also includes the data from the California rangeland compost experiment.

\subsection*{Funding}
 
JVS, PBS, and EWS report no individual support. Generous support for this project was provided by the California Climate Change 4th Assessment, the Rathmann Family Foundation, the US Department of Energy project with Sandia National Lab Project \# 24-233064 and Umakant Mishra, and USDA's McIntire Stennis to WLS.

\subsection*{Data availability}

All code used to implement our estimators, data analyses, and simulations are available as R scripts on Github (\url{https://github.com/spertus/SOC-storage-potential}). The Github also includes the data from the California rangeland compost experiment at \texttt{Data/silver\_lab\_nrcs\_data.csv}.

\bibliography{soil_bib}

@article{amelung1999minimisation,
  author    = {Amelung, W. and Zech, W.},
  title     = {Minimisation of organic matter disruption during particle-size fractionation of grassland epipedons},
  journal   = {Geoderma},
  volume    = {92},
  number    = {1},
  pages     = {73--85},
  year      = {1999},
  doi       = {10.1016/S0016-7061(99)00023-3}
}

@article{athey2021policy,
  author    = {Athey, Susan and Wager, Stefan},
  title     = {Policy learning with observational data},
  journal   = {Econometrica},
  volume    = {89},
  number    = {1},
  pages     = {133--161},
  year      = {2021},
  doi       = {10.3982/ECTA15732}
}

@article{andrews2024inference,
  author    = {Andrews, Isaiah and Kitagawa, Toru and McCloskey, Adam},
  title     = {Inference on winners},
  journal   = {The Quarterly Journal of Economics},
  volume    = {139},
  number    = {1},
  pages     = {305--358},
  year      = {2024},
  doi       = {10.1093/qje/qjad043}
}

@article{lal2004,
author = {R. Lal },
title = {Soil Carbon Sequestration Impacts on Global Climate Change and Food Security},
journal = {Science},
volume = {304},
number = {5677},
pages = {1623-1627},
year = {2004},
doi = {10.1126/science.1097396},
URL = {https://www.science.org/doi/abs/10.1126/science.1097396},
eprint = {https://www.science.org/doi/pdf/10.1126/science.1097396}}

@book{berry2004unsettling,
  author    = {Berry, W.},
  title     = {The Unsettling of America: Culture and Agriculture},
  edition   = {Revised},
  address   = {San Francisco},
  publisher = {Counterpoint},
  year      = {2004}
}

@article{bossio2020role,
  author    = {Bossio, D. A. and Cook-Patton, S. C. and Ellis, P. W. and Fargione, J. and Sanderman, J. and Smith, P. and Wood, S. and others},
  title     = {The role of soil carbon in natural climate solutions},
  journal   = {Nature Sustainability},
  volume    = {3},
  number    = {5},
  pages     = {391--398},
  year      = {2020},
  doi       = {10.1038/s41893-020-0491-z}
}

@article{bybeefinley2024deriving,
  author    = {Bybee-Finley, K. Ann and Muller, Katherine and White, Kathryn E. and Bowles, Timothy M. and Cavigelli, Michel A. and Han, Eunjin and Schomberg, Harry H. and Snapp, Sieglinde and Viens, Frederi},
  title     = {Deriving general principles of agroecosystem multifunctionality with the Diverse Rotations Improve Valuable Ecosystem Services (DRIVES) Network},
  journal   = {Agronomy Journal},
  volume    = {116},
  number    = {6},
  pages     = {2934--2951},
  year      = {2024},
  doi       = {10.1002/agj2.21697}
}

@article{chenu2019increasing,
  author    = {Chenu, C. and Angers, D. A. and Barr{\'e}, P. and Derrien, D. and Arrouays, D. and Balesdent, J.},
  title     = {Increasing organic stocks in agricultural soils: Knowledge gaps and potential innovations},
  journal   = {Soil and Tillage Research},
  volume    = {188},
  pages     = {41--52},
  year      = {2019},
  month     = {May},
  note      = {Soil Carbon and Climate Change: The 4 per Mille Initiative},
  doi       = {10.1016/j.still.2018.04.011}
}

@article{craig2021biological,
  author    = {Craig, M. E. and Mayes, M. A. and Sulman, B. N. and Walker, A. P.},
  title     = {Biological mechanisms may contribute to soil carbon saturation patterns},
  journal   = {Global Change Biology},
  volume    = {27},
  pages     = {2633--2644},
  year      = {2021},
  doi       = {10.1111/gcb.15584}
}

@article{curtis1964estimating,
  author    = {Curtis, R. O. and Post, B. W.},
  title     = {Estimating bulk density from organic-matter content in some {Vermont} forest soils},
  journal   = {Soil Science Society of America Journal},
  volume    = {28},
  pages     = {285--286},
  year      = {1964},
  doi       = {10.2136/sssaj1964.03615995002800020044x}
}

@article{delonge2013lifecycle,
  author    = {DeLonge, M. S. and Ryals, R. and Silver, W. L.},
  title     = {A lifecycle model to evaluate carbon sequestration potential and greenhouse gas dynamics of managed grasslands},
  journal   = {Ecosystems},
  volume    = {16},
  pages     = {962--979},
  year      = {2013},
  doi       = {10.1007/s10021-013-9660-5}
}

@article{ding2019decomposing,
  author    = {Ding, P. and Feller, A. and Miratrix, L.},
  title     = {Decomposing treatment effect variation},
  journal   = {Journal of the American Statistical Association},
  volume    = {114},
  number    = {525},
  pages     = {304--317},
  year      = {2019},
  doi       = {10.1080/01621459.2017.1407322}
}

@article{egami2023elements,
  author    = {Egami, N. and Hartman, E.},
  title     = {Elements of external validity: Framework, design, and analysis},
  journal   = {American Political Science Review},
  volume    = {117},
  number    = {3},
  pages     = {1070--1088},
  year      = {2023},
  doi       = {10.1017/S0003055422000880}
}

@book{fisher1925statistical,
  author    = {Fisher, R. A.},
  title     = {Statistical Methods for Research Workers},
  address   = {Edinburgh},
  publisher = {Oliver and Boyd},
  year      = {1925}
}

@article{georgiou2025soil,
  author    = {Georgiou, K. and Angers, D. and Champiny, R. E. and Cotrufo, M. F. and Craig, M. E. and Doetterl, S. and Grandy, A. S. and Lavallee, J. M. and Lin, Y. and Lugato, E. and Poeplau, C. and Rocci, K. S. and Schweizer, S. A. and Six, J. and Wieder, W. R.},
  title     = {Soil carbon saturation: What do we really know?},
  journal   = {Global Change Biology},
  volume    = {31},
  pages     = {e70197},
  year      = {2025},
  doi       = {10.1111/gcb.70197}
}

@article{gruijter2016farm,
  author    = {de Gruijter, J. J. and McBratney, A. B. and Minasny, B. and Wheeler, I. and Malone, B. P. and Stockmann, U.},
  title     = {Farm-scale soil carbon auditing},
  journal   = {Geoderma},
  volume    = {265},
  pages     = {120--130},
  year      = {2016},
  doi       = {10.1016/j.geoderma.2015.11.010}
}

@article{hassink1997capacity,
  author    = {Hassink, J.},
  title     = {The capacity of soils to preserve organic {C} and {N} by their association with clay and silt particles},
  journal   = {Plant and Soil},
  volume    = {191},
  number    = {1},
  pages     = {77--87},
  year      = {1997},
  doi       = {10.1023/A:1004213929699}
}

@book{hsu1996multiple,
  author    = {Hsu, J.},
  title     = {Multiple Comparisons: Theory and Methods},
  address   = {London},
  publisher = {Chapman and Hall},
  year      = {1996}
}

@article{hurlbert1984pseudoreplication,
  author    = {Hurlbert, S. H.},
  title     = {Pseudoreplication and the design of ecological field experiments},
  journal   = {Ecological Monographs},
  volume    = {54},
  number    = {2},
  pages     = {187--211},
  year      = {1984},
  doi       = {10.2307/1942661}
}

@book{imbens2015causal,
  author    = {Imbens, G. W. and Rubin, D. B.},
  title     = {Causal Inference for Statistics, Social, and Biomedical Sciences: An Introduction},
  address   = {Cambridge},
  publisher = {Cambridge University Press},
  year      = {2015},
  doi       = {10.1017/CBO9781139025751}
}

@misc{iea2023netzero,
  author    = {{International Energy Agency}},
  title     = {Net Zero Roadmap: A Global Pathway to Keep the 1.5 {\textdegree}{C} Goal in Reach -- Analysis},
  year      = {2023},
  publisher = {IEA},
  url       = {https://www.iea.org/reports/net-zero-roadmap-a-global-pathway-to-keep-the-15-0c-goal-in-reach}
}

@article{kitagawa2018who,
  author    = {Kitagawa, T. and Tetenov, A.},
  title     = {Who should be treated? {E}mpirical welfare maximization methods for treatment choice},
  journal   = {Econometrica},
  volume    = {86},
  number    = {2},
  pages     = {591--616},
  year      = {2018},
  doi       = {10.3982/ECTA13288}
}

@article{kravchenko2011whole,
  author    = {Kravchenko, A. N. and Robertson, G. P.},
  title     = {Whole-profile soil carbon stocks: The danger of assuming too much from analyses of too little},
  journal   = {Soil Science Society of America Journal},
  volume    = {75},
  number    = {1},
  pages     = {235--240},
  year      = {2011}
}

@article{kunzel2019metalearners,
  author    = {K{\"u}nzel, S. R. and Sekhon, J. S. and Bickel, P. J. and Yu, B.},
  title     = {Metalearners for estimating heterogeneous treatment effects using machine learning},
  journal   = {Proceedings of the National Academy of Sciences},
  volume    = {116},
  number    = {10},
  pages     = {4156--4165},
  year      = {2019},
  doi       = {10.1073/pnas.1804597116}
}

@article{mayer2026cost,
  author    = {Mayer, A. and Dumortier, J. and Hausfather, Z. and Pett-Ridge, J. and Slessarev, E. W.},
  title     = {The value of reversible carbon storage in a zero-emissions world},
  journal   = {Environmental Science \& Technology},
  volume    = {60},
  number    = {24},
  year      = {2026},
  doi       = {10.1021/acs.est.6c00333}
}

@article{mayer2022climate,
  author    = {Mayer, A. and Silver, W. L.},
  title     = {The climate change mitigation potential of annual grasslands under future climates},
  journal   = {Ecological Applications},
  volume    = {32},
  number    = {8},
  year      = {2022},
  doi       = {10.1002/eap.2705}
}

@article{minami1994methane,
  author    = {Minami, K.},
  title     = {Methane from rice production},
  journal   = {Fertilizer Research},
  volume    = {37},
  number    = {3},
  pages     = {167--179},
  year      = {1994},
  doi       = {10.1007/BF00748935}
}

@article{neyman1923sur,
  author    = {Neyman, J.},
  title     = {Sur les applications de la th{\'e}orie des probabilit{\'e}s aux experiences agricoles: Essai des principes},
  journal   = {Roczniki Nauk Rolniczki},
  volume    = {10},
  pages     = {1--51},
  year      = {1923}
}

@misc{nitta2022rice,
  author    = {Nitta, N.},
  title     = {Rice farming to restore soil},
  year      = {2022},
  url       = {https://www.earthisland.org/journal/index.php/articles/entry/rice-farming-to-restore-soil/}
}

@article{oldfield2022crediting,
  author    = {Oldfield, E. E. and Eagle, A. J. and Rubin, R. R. and Rudek, J. and Sanderman, J. and Gordon, D. R.},
  title     = {Crediting agricultural soil carbon sequestration},
  journal   = {Science},
  volume    = {375},
  number    = {6586},
  pages     = {1222--1225},
  year      = {2022},
  doi       = {10.1126/science.abl7991}
}

@article{plevin2010greenhouse,
  author    = {Plevin, R. J. and O'Hare, M. and Jones, A. D. and Torn, M. S. and Gibbs, H. K.},
  title     = {Greenhouse gas emissions from biofuels' indirect land use change are uncertain but may be much greater than previously estimated},
  journal   = {Environmental Science \& Technology},
  volume    = {44},
  number    = {21},
  pages     = {8015--8021},
  year      = {2010},
  doi       = {10.1021/es101946t}
}

@unpublished{polussa2026toward,
  author    = {Polussa, A. and Oldfield, E. E. and Runge, T. and Evans, E. and Liu, S. S. and Forbes, E. and Partida, C. and Potash, E. and Sanderman, J. and Sheffer, M. and Bradford, M. A.},
  title     = {Toward causal designs to quantify management-driven soil carbon change at multi-field scales},
  note      = {agriRxiv preprint},
  year      = {2026},
  url       = {https://www.cabidigitallibrary.org/doi/pdf/10.31220/agriRxiv.2026.00422}
}

@article{potash2023multi,
  author    = {Potash, E. and Guan, K. and Margenot, A. J. and Lee, D. K. and Boe, A. and Douglass, M. and Heaton, E. and others},
  title     = {Multi-site evaluation of stratified and balanced sampling of soil organic carbon stocks in agricultural fields},
  journal   = {Geoderma},
  volume    = {438},
  pages     = {116587},
  year      = {2023},
  doi       = {10.1016/j.geoderma.2023.116587}
}

@article{qian2011performance,
  author    = {Qian, Min and Murphy, Susan A.},
  title     = {Performance guarantees for individualized treatment rules},
  journal   = {The Annals of Statistics},
  volume    = {39},
  number    = {2},
  pages     = {1180--1210},
  year      = {2011},
  doi       = {10.1214/10-AOS864}
}

@book{rosenbaum2002observational,
  author    = {Rosenbaum, P. R.},
  title     = {Observational Studies},
  address   = {New York, NY},
  publisher = {Springer},
  year      = {2002},
  url       = {https://link.springer.com/book/10.1007/978-1-4757-3692-2}
}

@article{sanford2012soil,
  author    = {Sanford, G. R. and Posner, J. L. and Jackson, R. D. and Kucharik, C. J. and Hedtcke, J. L. and Lin, T.},
  title     = {Soil carbon lost from {M}ollisols of the {N}orth {C}entral {U.S.A.} with 20 years of agricultural best management practices},
  journal   = {Agriculture, Ecosystems \& Environment},
  volume    = {162},
  pages     = {68--76},
  year      = {2012},
  month     = {November},
  doi       = {10.1016/j.agee.2012.08.011}
}

@article{schrumpf2013storage,
  author    = {Schrumpf, M. and Kaiser, K. and Guggenberger, G. and Persson, T. and K{\"o}gel-Knabner, I. and Schulze, E.-D.},
  title     = {Storage and stability of organic carbon in soils as related to depth, occlusion within aggregates, and attachment to minerals},
  journal   = {Biogeosciences},
  volume    = {10},
  pages     = {1675--1691},
  year      = {2013},
  doi       = {10.5194/bg-10-1675-2013}
}

@book{scott1999seeing,
  author    = {Scott, J. C.},
  title     = {Seeing Like a State: How Certain Schemes to Improve the Human Condition Have Failed},
  address   = {New Haven, CT and London},
  publisher = {Yale University Press},
  year      = {1999}
}

@techreport{silver2018carbon,
  author      = {Silver, W. and Vergara, S. and Mayer, A.},
  title       = {Carbon sequestration and greenhouse gas mitigation potential of composting and soil amendments on {California's} rangelands},
  institution = {California's Fourth Climate Change Assessment},
  year        = {2018},
  url         = {https://www.energy.ca.gov/media/2059}
}

@article{six2024six,
  author    = {Six, J. and Doetterl, S. and Laub, M. and M{\"u}ller, C. R. and Van de Broek, M.},
  title     = {The six rights of how and when to test for soil {C} saturation},
  journal   = {SOIL},
  volume    = {10},
  number    = {1},
  pages     = {275--279},
  year      = {2024},
  doi       = {10.5194/soil-10-275-2024}
}

@article{slessarev2023initial,
  author    = {Slessarev, E. W. and Mayer, A. and Kelly, C. and Georgiou, K. and Pett-Ridge, J. and Nuccio, E. E.},
  title     = {Initial soil organic carbon stocks govern changes in soil carbon: Reality or artifact?},
  journal   = {Global Change Biology},
  volume    = {29},
  number    = {5},
  pages     = {1239--1247},
  year      = {2023},
  doi       = {10.1111/gcb.16491}
}

@article{smith2005overview,
  author    = {Smith, P.},
  title     = {An overview of the permanence of soil organic carbon stocks: Influence of direct human-induced, indirect and natural effects},
  journal   = {European Journal of Soil Science},
  volume    = {56},
  number    = {5},
  pages     = {673--680},
  year      = {2005},
  doi       = {10.1111/j.1365-2389.2005.00708.x}
}

@article{spertus2021optimal,
  author    = {Spertus, J. V.},
  title     = {Optimal sampling and assay for estimating soil organic carbon},
  journal   = {Open Journal of Soil Science},
  volume    = {11},
  pages     = {93--121},
  year      = {2021},
  doi       = {10.4236/ojss.2021.112006}
}

@article{stanley2023valid,
  author    = {Stanley, P. and Spertus, J. V. and Chiartas, J. and Stark, P. B. and Bowles, T.},
  title     = {Valid inferences about soil carbon in heterogeneous landscapes},
  journal   = {Geoderma},
  volume    = {430},
  pages     = {116323},
  year      = {2023},
  month     = {February},
  doi       = {10.1016/j.geoderma.2022.116323}
}

@article{stewart2007soil,
  author    = {Stewart, C. E. and Paustian, K. and Conant, R. T. and Plante, A. F. and Six, J.},
  title     = {Soil carbon saturation: Concept, evidence and evaluation},
  journal   = {Biogeochemistry},
  volume    = {86},
  number    = {1},
  pages     = {19--31},
  year      = {2007},
  doi       = {10.1007/s10533-007-9140-0}
}

@article{thamo2016challenges,
  author    = {Thamo, T. and Pannell, D. J.},
  title     = {Challenges in developing effective policy for soil carbon sequestration: Perspectives on additionality, leakage, and permanence},
  journal   = {Climate Policy},
  volume    = {16},
  number    = {8},
  pages     = {973--992},
  year      = {2016},
  doi       = {10.1080/14693062.2015.1075372}
}

@article{viscarrarossel2024how,
  author    = {Viscarra Rossel, R. A. and Webster, R. and Zhang, M. and Shen, Z. and Dixon, K. and Wang, Y.-P. and Walden, L.},
  title     = {How much organic carbon could the soil store? {T}he carbon sequestration potential of {Australian} soil},
  journal   = {Global Change Biology},
  volume    = {30},
  number    = {1},
  pages     = {e17053},
  year      = {2024},
  doi       = {10.1111/gcb.17053}
}

@article{wendt2013equivalent,
  author    = {Wendt, J. W. and Hauser, S.},
  title     = {An equivalent soil mass procedure for monitoring soil organic carbon in multiple soil layers},
  journal   = {European Journal of Soil Science},
  volume    = {64},
  number    = {1},
  pages     = {58--65},
  year      = {2013},
  doi       = {10.1111/ejss.12002}
}

@article{wuest2024temporal,
  author    = {Wuest, Stewart B. and Durfee, Nicole},
  title     = {Temporal variability is a major source of uncertainty in soil carbon measurements},
  journal   = {Soil Science Society of America Journal},
  volume    = {88},
  number    = {3},
  pages     = {830--845},
  year      = {2024}
}

\newpage

 \appendix

\renewcommand\thefigure{S\arabic{figure}}    

\setcounter{figure}{0}    
 
 \section*{Supplementary Material}

\begin{figure}[h]
    \centering
     \includegraphics[width = 0.9 \textwidth]{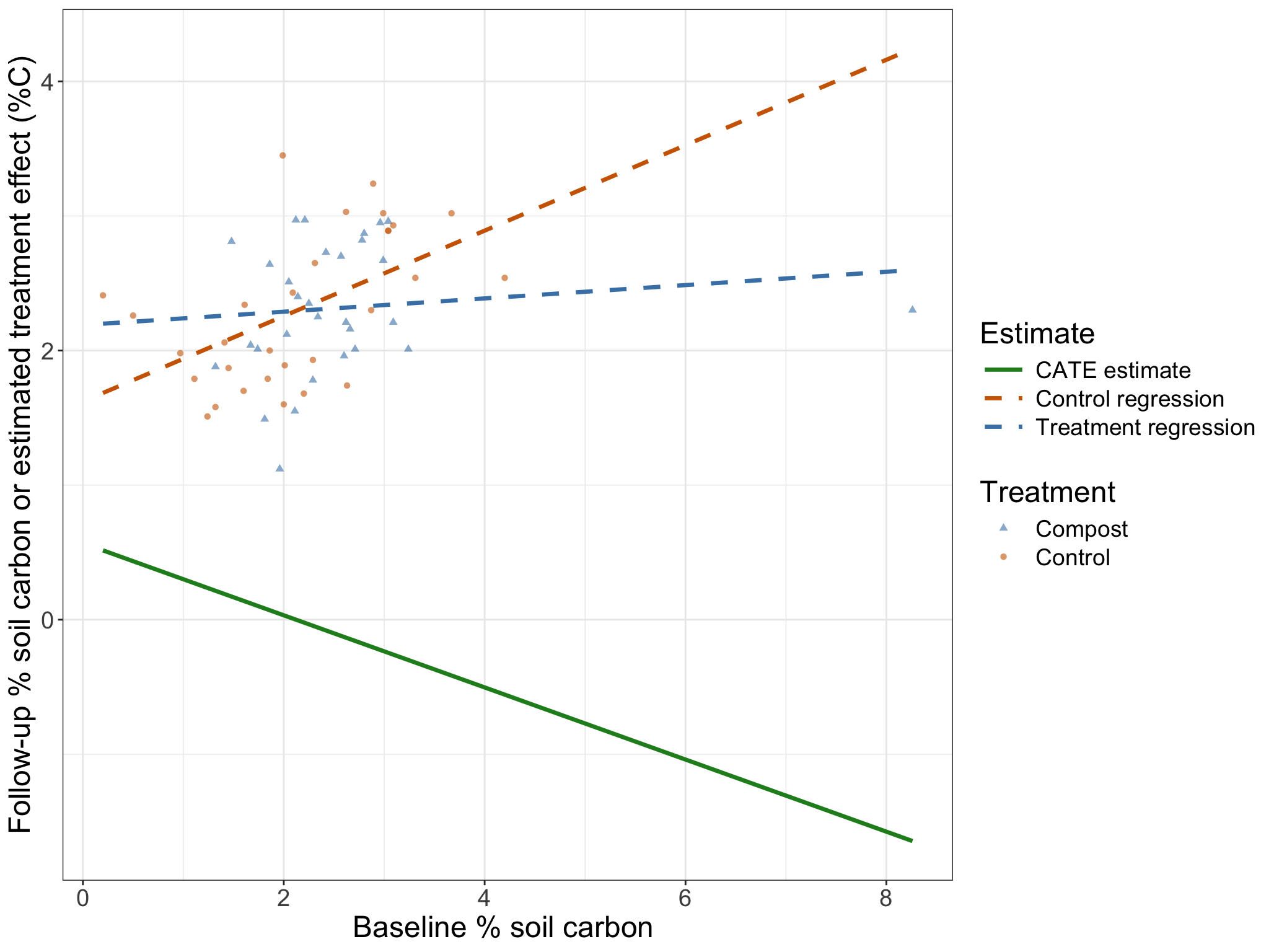}
    \caption{Illustration of estimating a conditional average treatment effect (CATE) curve from raw treatment data (blue triangles) and control data (orange circles). The CATE in question is the expected treatment effect on follow-up \% soil carbon (y-axis) at any given level of baseline \% soil carbon (x-axis). Ordinary least squares is used to fit a regression in the treatment group (dashed blue line) and a regression in the control group (dashed orange line). The CATE estimate (green line) is the dashed blue (treatment) line minus the dashed orange (control) line. The data come from an experiment evaluating the effects of compost amendments on California rangelands.}
    \label{fig:appendix_cate_intuition}
\end{figure}

\begin{figure}[p]
    \centering
     \includegraphics[width = 0.9 \textwidth]{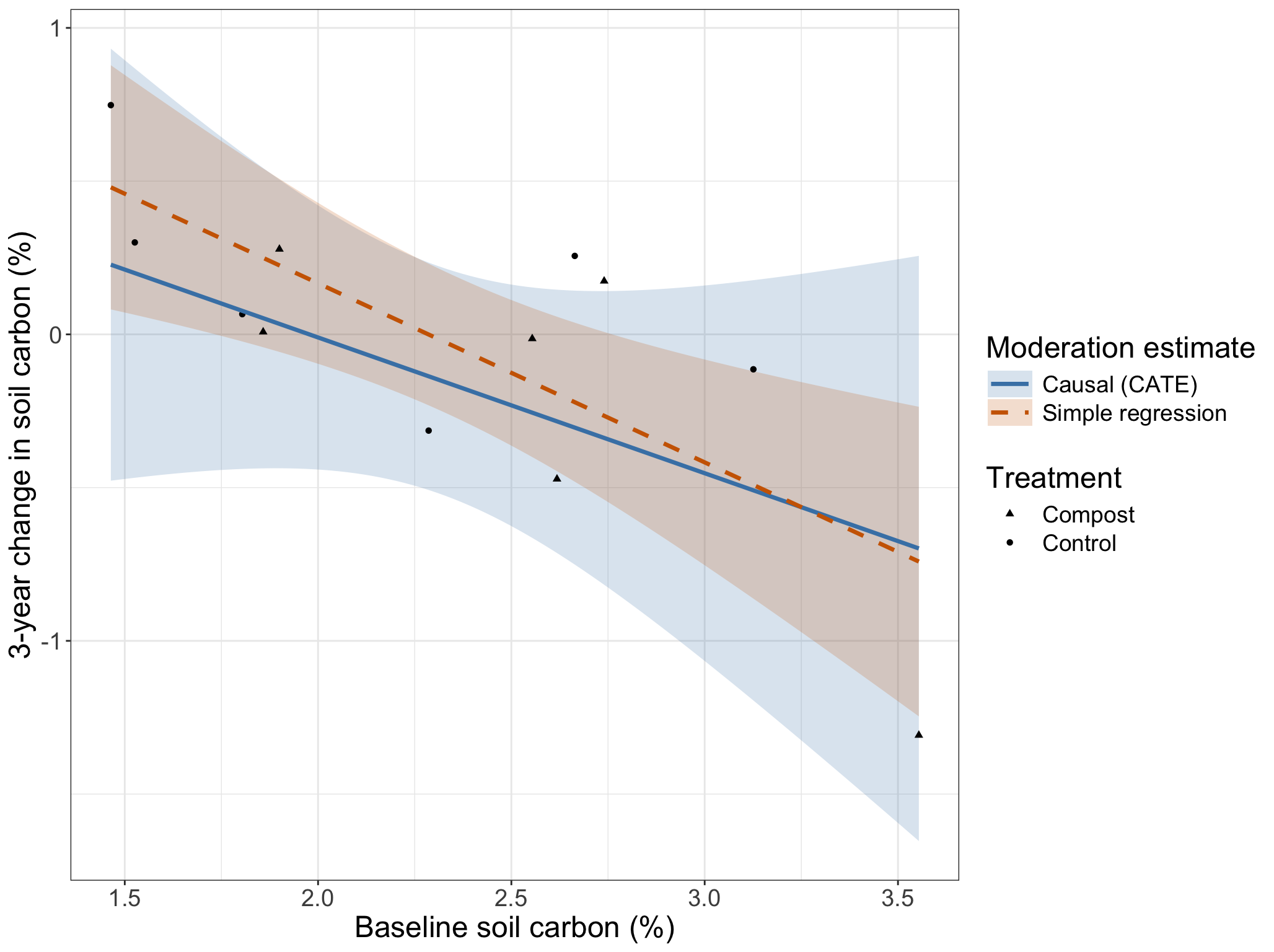}
    \caption{Estimates of conditional average treatment effect (CATE; green line) and regression of change in \%C on baseline \%C after aggregating data to the plot-level by averaging core-level estimates within plots. }
    \label{fig:appendix_saturation_plotlevel}
\end{figure}

\begin{figure}[p]
    \centering
     \includegraphics[width = 0.9 \textwidth]{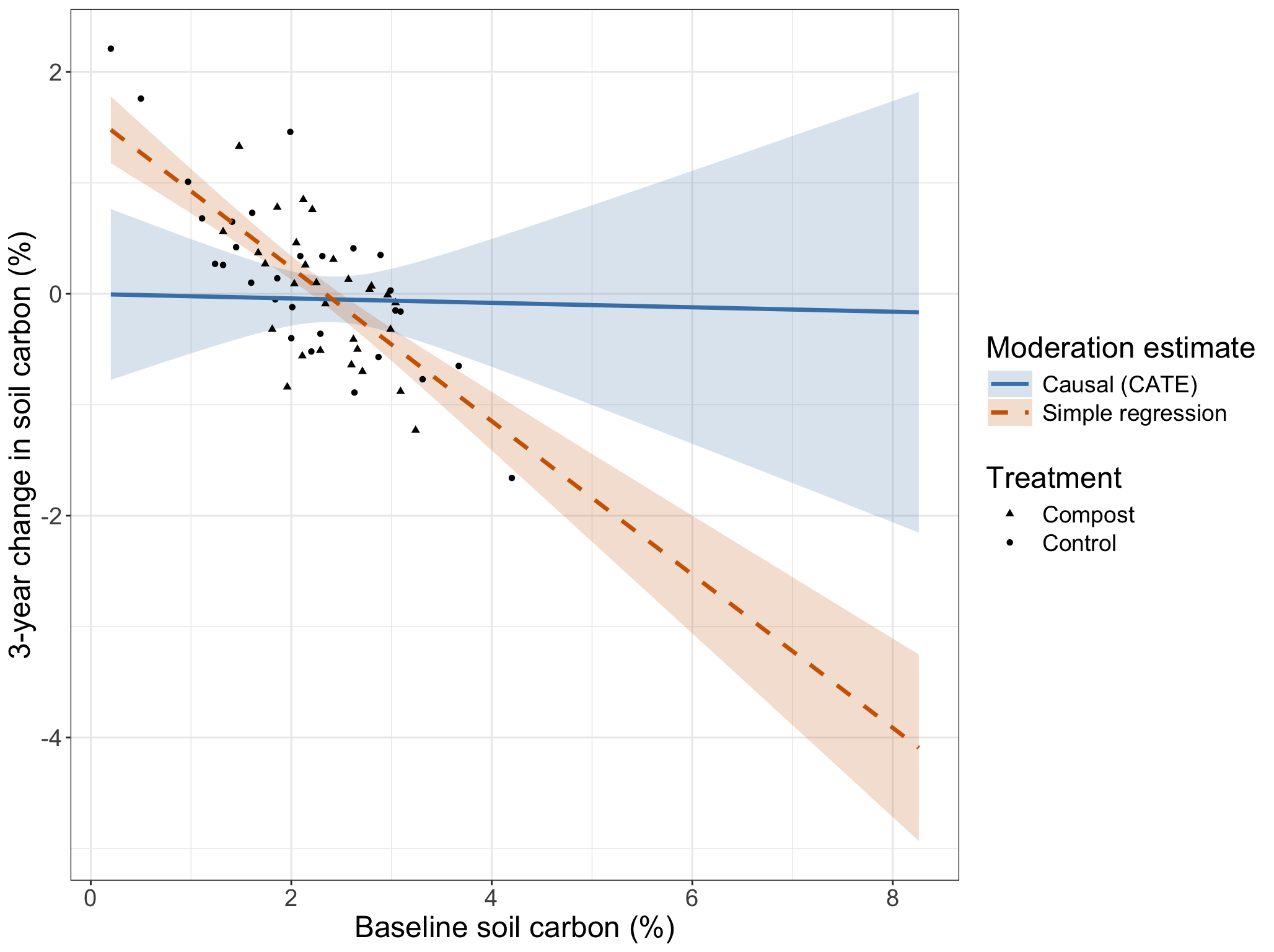}
    \caption{Plot of conditional average treatment effect (CATE; green line) and regression of 3-year change in \%C on baseline \%C, removing the outlying point at ~8\% baseline C. The CATE estimate is close to 0, indicating no treatment effect at any level of baseline \%C.}
    \label{fig:appendix_saturation_outlierremoved}
\end{figure}

\end{document}